\documentstyle[aps,epsfig,twocolumn,prl]{revtex}
\input epsf
\def\cassg {\mbox{$\Xi^\circ\rightarrow\Sigma^\circ\gamma $}}

\def\caslpi {\mbox{$\Xi^\circ\rightarrow\Lambda\pi^\circ$}}
\def\siglg {\mbox{$\Sigma^\circ\rightarrow\Lambda\gamma$}}
\def\lamppi {\mbox{$\Lambda\rightarrow p\pi^-$}}

\begin{document}                          

\draft

\title{
A Measurement of the Branching Ratio and Asymmetry 
of the Decay $\cassg$}


\parindent=0.in
\parskip 0 in

\author {
A.~Alavi-Harati$^{12}$,
T.~Alexopoulos$^{12,\star}$,
M.~Arenton$^{11}$,
K.~Arisaka$^2$,
S.~Averitte$^{10}$,
A.R.~Barker$^5$,
L.~Bellantoni$^7$,
A.~Bellavance$^9$,
J.~Belz$^{10,\ddagger}$,
R.~Ben-David$^{7}$,
D.R.~Bergman$^{10}$,
E.~Blucher$^4$, 
G.J.~Bock$^7$,
C.~Bown$^4$, 
S.~Bright$^4$,
E.~Cheu$^1$,
S.~Childress$^7$,
R.~Coleman$^7$,
M.D.~Corcoran$^9$,
G.~Corti$^{11}$, 
B.~Cox$^{11}$,
M.B.~Crisler$^7$,
A.R.~Erwin$^{12}$,
R.~Ford$^7$,
A.~Glazov$^4$,
A.~Golossanov$^{11}$,
G.~Graham$^{4}$, 
J.~Graham$^4$,
K.~Hagan$^{11}$,
E.~Halkiadakis$^{10}$,
J.~Hamm$^1$,
K.~Hanagaki$^{8,\bullet}$,  
S.~Hidaka$^8$,
Y.B.~Hsiung$^7$,
V.~Jejer$^{11}$,
D.A.~Jensen$^7$,
R.~Kessler$^4$,
H.G.E.~Kobrak$^{3}$,
J.~LaDue$^5$,
A.~Lath$^{10}$,
A.~Ledovskoy$^{11}$,
P.L.~McBride$^7$,
P.~Mikelsons$^5$,
E.~Monnier$^{4,*}$,
T.~Nakaya$^{7,\parallel}$,
K.S.~Nelson$^{11}$,
H.~Nguyen$^7$,
V.~O'Dell$^7$, 
M.~Pang$^7$, 
R.~Pordes$^7$,
V.~Prasad$^4$, 
B.~Quinn$^4$,
E.J.~Ramberg$^{7,\dagger}$, 
R.E.~Ray$^7$,
A.~Roodman$^{4,\&}$, 
M.~Sadamoto$^8$, 
S.~Schnetzer$^{10}$,
K.~Senyo$^{8,\#}$, 
P.~Shanahan$^7$,
P.S.~Shawhan$^{4,\P}$,
J.~Shields$^{11}$,
W.~Slater$^2$,
N.~Solomey$^4$,
S.V.~Somalwar$^{10}$, 
R.L.~Stone$^{10}$, 
I.~Suzuki$^{8,\S}$,
E.C.~Swallow$^{4,6}$,
S.A.~Taegar$^1$,
R.J.~Tesarek$^{10,\S}$, 
G.B.~Thomson$^{10}$,
P.A.~Toale$^5$,
A.~Tripathi$^2$,
R.~Tschirhart$^7$,
S.E.~Turner$^2$ 
Y.W.~Wah$^4$,
J.~Wang$^1$,
H.B.~White$^7$, 
J.~Whitmore$^7$,
B.~Winstein$^4$, 
R.~Winston$^4$, 
T.~Yamanaka$^8$,
E.D.~Zimmerman$^{4,\pounds}$
\vspace*{.1 in} 
\footnotesize
$^1$ University of Arizona, Tucson, Arizona 85721 \\
$^2$ University of California at Los Angeles, Los Angeles, California 90095 \\
$^{3}$ University of California at San Diego, La Jolla, California 92093 \\
$^4$ The Enrico Fermi Institute, The University of Chicago, 
Chicago, Illinois 60637 \\
$^5$ University of Colorado, Boulder, Colorado 80309 \\
$^6$ Elmhurst College, Elmhurst, Illinois 60126 \\
$^7$ Fermi National Accelerator Laboratory, Batavia, Illinois 60510 \\
$^8$ Osaka University, Toyonaka, Osaka 560-0043 Japan \\
$^9$ Rice University, Houston, Texas 77005 \\
$^{10}$ Rutgers University, Piscataway, New Jersey 08854 \\
$^{11}$ The Department of Physics and Institute of Nuclear and 
Particle Physics, University of Virginia, 
Charlottesville, Virginia 22901 \\
$^{12}$ University of Wisconsin, Madison, Wisconsin 53706 \\
$^{\dagger}$ To whom correspondence should be addressed. \\
$^{*}$ On leave from C.P.P. Marseille/C.N.R.S., France \\
$^{\star}$ Current address, National Technical University, 175 73, Athens,
Greece\\
$^\ddagger$ Current address, Montana State University, Bozeman, Montana
59717\\
$^{\bullet}$ Current address, Princeton University, Princeton, 
New Jersey 08544\\
$^{\parallel}$ Current address, Kyoto University, 606-8502 Japan\\
$^{\&}$ Current address Stanford Linear Accelerator Center, Stanford,
California 94309 \\
$^{\#}$ Current address, Nagoya University, Nagoya 464-8602 Japan \\
$^\P$ Current address, California Institute of Technology, Pasadena, 
California 91125 \\
$^\S$ Current address Fermi National Accelerator Laboratory, Batavia, 
 Illinois 60510 \\
$^{\pounds}$ Current address, Columbia University, New York, New York 10027\\
\vspace*{.1 in}
\centerline{ \bf The KTeV Collaboration}
}

\normalsize

\maketitle
 
\begin{abstract}
We have studied the rare weak radiative hyperon decay
$\cassg$ in the KTeV experiment at Fermilab.  We have identified 4045 signal 
events over a background of 804 events.  
The dominant $\caslpi$ decay, which was used for
normalization, is the only important background source.
An analysis of the acceptance of
both modes yields a branching ratio of 
$BR(\cassg)/BR(\caslpi)=(3.34\pm0.05\pm0.09)\times 10^{-3}$.  
By analyzing the final state
decay distributions, we have also determined 
that the $\Sigma^\circ$ emission asymmetry 
parameter for this decay is $\alpha_{\Xi\Sigma} = -0.63\pm 0.09$.
\end{abstract}


We report here
on the Fermilab KTeV experiment's analysis of the decay $\cassg$.
This is an example of one of the most intriguing classes of baryon
decays--
weak radiative hyperon decays (WRHD).  
These decays are 
experimentally quite accessible, 
having branching ratios on the order
of $10^{-3}$\cite{PDG}.
The two most significant variables 
that can be measured in these decays are the branching ratio (BR), and
the asymmetry of the baryon emission with respect to the initial
spin axis ($\alpha$).  However, despite this experimental
simplicity, there is no known theoretical framework for explaining
these decays. Some of the difficulty in explaining
WRHD is because predictions using the quark model do not
match predictions from an analysis at the hadron level\cite{ZenLach}.

Hara proved in 1964
that $\alpha=0$ in the SU(3) limit for the 
$\Sigma^+$ and $\Xi^-$ WRHD, assuming only CP invariance
and U spin symmetry (s and d quark exchange symmetry)\cite{Hara}.
An estimate, based on single-quark $s\rightarrow d$ transitions and that takes 
into account SU(3) breaking predicts a modest positive asymmetry,
with the magnitude depending on the assignment of 
quark masses\cite{Vasanti}.
However, the only asymmetry of any WRHD
that has been measured accurately, the one for 
$\Sigma^+\rightarrow p\gamma$, has been found to be highly
negative, $-0.76\pm.08$\cite{E761}.  The corresponding branching ratio cannot
be understood as due to single-quark transition. 
A recent paper suggests that this large
negative asymmetry could result from intermediate resonances in the decay, and
gives a prediction of a positive asymmetry for the 
$\cassg$ mode\cite{Holstein}.

To constrain the theoretical models, 
it is vital to measure accurately the parameters of
WRHD for hyperons other than the $\Sigma^+$.
Reference \cite{Zen2000} states that whether 
Hara's theorem can be incorporated into a succesful
theoretical model depends
on the experimental results from the 
two $\Xi^\circ$ weak radiative decays.

The decay $\cassg$
has been previously observed in two separate experiments\cite{Teige,NA48}.
Both experiments have sample sizes of less than 100 events.  The most
accurate branching ratio is reported as 
$(3.6\pm0.4)\times 10^{-3}$\cite{Teige}.
This experiment also reported an asymmetry 
measurement, but it is incorrect because
the depolarization in the decay $\siglg$ was not properly taken into account
as discussed below.

Many of the details of the KTeV 
beamline, experiment and hyperon trigger can be
found in a previous publication\cite{beta}.  We reemphasize here some of
the details particularly relevant to this measurement.  
The beamline for KTeV was designed 
to deliver two square high intensity beams of $K_L$
particles for CP violation studies.  
The detector is situated 94 
meters from this target.  
The sweeping magnets in the beamline were designed and operated
so that the integrated magnetic field delivered $\Xi^\circ$ hyperons 
polarized (with about 10\% polarization) 
in the positive or negative vertical direction.  We reversed one
of the magnets regularly so that the net polarization was zero for the data
discussed here.

The detector consists of a 64 m long vacuum decay vessel followed by
a spectrometer.  The decay vessel contains several 
lead/scintillator vetos with
square apertures in their center.  These 
veto events where decay particles leave
the sensitive area of the detector.  The vacuum vessel is followed by
4 drift chambers, two situated on either side of an analysis magnet.  
This magnetic spectrometer gives
a momentum resolution for charged particles of 
$\sigma(P)/P=0.38\%\oplus P\times0.016\%$, where P is in GeV/c.  
Photon
veto detectors surround each drift chamber station.

Downstream of 
the last drift chamber is a set of 9 transition radiation detectors (TRD).
A plane of vertically oriented trigger 
hodoscopes is situated after the last TRD
chamber.  This plane is followed by the
electromagnetic calorimeter which is composed of an array of 
3100 crystals of pure CsI, 27 radiation lengths deep.  
The energy resolution of the calorimeter for electrons and photons is
approximately $\sigma(E)/E \approx 0.45\%\oplus2\%/\sqrt{E}$ 
and the position resolution is
approximately 1 mm.

There are
two beam holes in the TRD radiators, the trigger counters and the calorimeter.
After the calorimeter is a 10 cm thick lead wall and then a 
hodoscope array (HA),
with a single beam hole encompassing both beams,
which provides a veto for hadronic showers.  
This is followed
by three steel walls and 
two hodoscope arrays, which provide a veto for muons.
Two small scintillation counters are situated
in the beam hole of the first steel wall to provide a trigger for 
highly forward going charged particles.

The final decay products of the $\cassg$ signal are a proton in one of
the beam holes
and  a $\pi^-$ and two photons  hitting the calorimeter.  These are
the same final decay products as for the normalization mode $\caslpi$.

The trigger for the experiment is built on a
standard multi-level architecture, with
the Level 1 incorporating fast triggering elements, Level 2 being a somewhat
slower and more detailed hardware trigger and Level 3 being a real time
processing filter to select for specific decay modes.  The main hyperon trigger
was designed to measure the previously unobserved
beta decay of the $\Xi^\circ$.  Further details can be found in
the paper reporting its discovery\cite{beta}.  
Since this trigger vetoed on a signal of more than 2.5
MIP's in the HA array, it rejected 
58\% of the $\cassg$ signal and $\caslpi$ normalization mode because the
final state $\pi^-$ often showers in the CsI or lead.  
This introduces no bias in the measurement of the branching ratio, however,
because of the identical final state between signal and 
normalization mode.

We established a set of offline selection cuts to accept both modes and another
set of cuts to distinguish between the two.  The cuts in common include the
following.  There are 2 charged tracks, with the higher momentum track
positively  charged and with momentum between 85 and 600 GeV/c.  The lower
momentum track is negative, with a momentum between 5 and 150 GeV/c and has
an  energy in the CsI calorimeter less than 90\% of its momentum.  
The two tracks must have a momentum ratio greater than 3.5 and, assuming
a proton and pion identity,
must combine to form an invariant mass within 15 MeV/c$^2$ of the $\Lambda$
mass.
Each of the two calorimeter clusters is required to have at least
2 GeV of energy and no charged track
pointing at it. The sum of the two energies must be 
at least 16 GeV.  We also require that the two photons do not
reconstruct as a $\pi^\circ$ within 3m along the beamline direction
of the charged track vertex. 

We reconstruct the $\Lambda$ flight path from the reconstructed momentum
of the proton and pion tracks.  To find the decay vertex of the parent
$\Xi^\circ$, we numerically vary the distance along the beamline,
or z.  For a $\Xi^\circ$ vertex at a given z, we calculate the transverse
coordinates by extrapolating the $\Lambda$ flight path 
back to that location.
We choose the value of z that minimizes the difference
between the reconstructed invariant mass of all the final state particles
and the $\Xi^\circ$ mass.  
We discard events with a decay vertex 
outside of the vacuum decay region,
which we define as Z=95-155 m.
Events are also discarded if the daughter $\Lambda$ vertex falls
upstream of the parent $\Xi^\circ$ vertex.
We also require that the total reconstructed momentum ($\Lambda$ plus
two photons)
transverse to the flight path of the $\Xi^\circ$ parent, or $p_T^2$, be
less than .0005 $GeV^2/c^2$.   In addition, we
reconstruct the event as a decay of a $\Lambda$ from the primary target
and as a
$\Xi^\circ\rightarrow\Lambda\gamma$ decay and discard the event if it
is consistent with either.

The distinguishing characteristics between the signal and normalization decay
modes reside in the invariant masses between pairs of the final state
particles.  The signal mode will have one of the $\Lambda\gamma$
combinations reconstruct as a $\Sigma^\circ$, while the normalization
mode will have a $\gamma\gamma$ reconstruction near the $\pi^\circ$ mass.
Figure 1 shows these invariant mass combinations for the data.  A cut of
$|m_{\gamma\gamma}-m_{\pi^\circ}|<20$ Mev/c$^2$ is made in this figure to
make the signal mode visible.  This cut gives the optimum
value for signal to background.  The lower
energy photon is chosen to be paired with the $\Lambda$.  The presence
of a $\Sigma^\circ$ mass band 
can be seen in this plot along with the band arising from the incorrect
pairing of the photon.
\vspace*{.1 in}

\begin{figure} 
\centerline{\psfig{figure=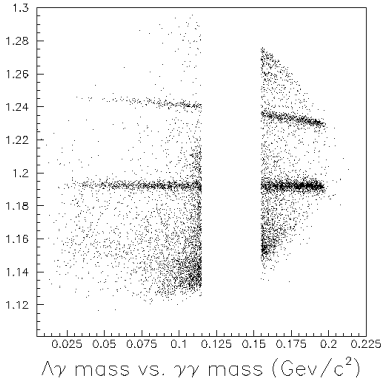,width=6cm}}
\caption{A plot of $\Lambda-\gamma$ mass versus $\gamma-\gamma$
mass for all data. }
\label{masses}
\end{figure}
\begin{figure} 
\centerline{\psfig{figure=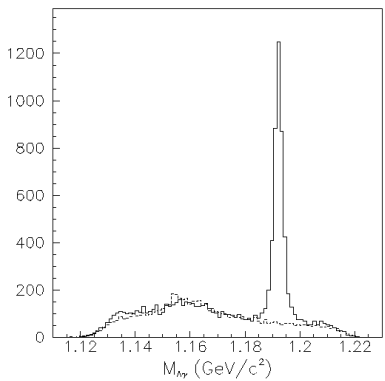,width=6cm}}
\caption{A plot of the $\Lambda-\gamma$ invariant mass for all data events
passing selection cuts.  Superimposed is normal mode Monte Carlo
events that pass the selection cuts.  This latter distribution is scaled
to fit the data.}
\label{peak}
\end{figure}

Figure 2 shows the final invariant $\Lambda\gamma$
mass plot (a projection of Figure 1), 
with the photon chosen to give the lowest mass.
The background under the  $\Sigma^\circ$ mass peak 
can be explained entirely by the
normalization mode decays $\caslpi$ that fall outside of our $\gamma\gamma$
mass cut.  
We fit the distribution in Figure 2 above a mass value of 1.16 GeV/c$^2$ 
to a combination of Monte Carlo simulations of the 
$\cassg$ signal and $\caslpi$ background.
The relative level of the background has 
been allowed to float in the fit since the
simulation is not guaranteed to be accurate in describing the population
at the extreme tails of
the $\pi^\circ$ mass peak. The fit level is about 25\% higher than indicated
by the number of well reconstructed normalization mode decays.

As can be seen in Figure 2, the simulated background describes the data
very well.
We subtract the simulated background distribution from the
data and count the number of 
events in the $\Sigma^\circ$ mass region of 1185-1201 MeV/c$^2$.  For our
data, this is 4045 events, which gives a statistical uncertainty
of $\pm 64$ events.  The number of background events in this region
as estimated by the background fit is $804$ events, giving a signal to
background ratio of 5:1. 
The number of normalization events in our data is 1377642.  There is a
negligible amount of background for the normalization mode.
Using the 
simulation of the detector, we determine that the ratio of acceptances
for the signal decay mode to the normalization decay mode is 0.88.
The acceptance does not depend significantly on the asymmetry parameter used.
This results in a branching ratio measurement of
$BR(\cassg)/BR(\caslpi)=(3.34\pm0.05)\times 10^{-3}$
where the error quoted is purely statistical.

The two largest systematic effects on the calculation of
the branching ratio are background and acceptance uncertainties.
The statistical error on the amount of background is 28.  We 
varied the region around the signal mass peak in which to fit the 
simulated background and obtained an uncertainty of 40 events in the 
signal region due to the level of the fit.  
Additionally, looking at comparisons between event variables for those 
background events in the $\Sigma^\circ$ mass peak sidebands, we can see small
discrepencies between the simulation and the data.  We conservatively 
estimate that at most 60 background events could be unaccounted
for under the signal peak.  In all, we assign an uncertainty of 80 events
to the amount of background, giving an uncertainty in the branching ratio
of $\pm 0.07\times 10^{-3}$.

The uncertainty in the acceptances is
primarily due to the fact that we do not simulate hadronic showering in the
HA detector and thus do not account for this effect at the trigger level.
The effect of the hadronic veto on the absolute level of our signal is
substantial.
However, because the normalization mode has the identical final
state, this effect cancels out in the branching ratio calculation.  
We find that the kinematic distributions relevant to the HA detector
(such as pion and proton position and momentum) are very similar between
the signal and normalization decay modes.
We tested a 
simple model of hadronic vetoing in the Monte Carlo containing both position
and momentum dependence for the $\pi^-$. We
observed that the ratio of acceptances between
signal and normalization did not change at all.
We assign a value of $\pm 0.05\times 10^{-3}$ branching ratio uncertainty 
due to the hadronic veto.

We explored the sensitivity to analysis cuts by varying the three most
relevant variables: $p_T^2$, the z range for the $\Xi^\circ$
and the separation of the $\Xi^\circ$ and $\Lambda$ vertices.
The variation seen from these cuts matches the level
of uncertainty from the other sources quoted above.
We also broke the data into various subsets (left vs right beam,
and run number) and find the individual results statistically consistent.

The final systematic uncertainty in the branching ratio from the above
sources is 
$\pm 0.09 \times 10^{-3}$.  This then gives the final branching ratio
result of $BR(\cassg)/BR(\caslpi)=(3.34\pm0.05\pm0.09)\times 10^{-3}$.

A value of the asymmetry parameter, $\alpha_{\Xi\Sigma}$, 
has also been obtained
from our data.  This parameter is determined most directly from the
up-down asymmetry in the decay distribution of polarized $\Xi^\circ$'s.
However, our low degree of polarization ($\approx 10\%$) limits the power
of this approach.  For an unpolarized initial $\Xi^\circ$ state, 
as in our total data
sample, angular momentum considerations dictate that the
longitudinal polarization of the decay $\Sigma^\circ$ is just 
$-\alpha_{\Xi\Sigma}$\cite{Behrends}.
In the subsequent electromagnetic decay, the $\Lambda$ retains 
the component of the parent $\Sigma^\circ$ polarization along the $\Lambda$
direction of motion, but with opposite sign.  The final $\lamppi$
weak decay asymmetry ($\alpha_{\Lambda p}=0.642$\cite{PDG}) then analyzes the 
polarization.  The resulting
proton angular distribution is thus proportional to 
$[1+\alpha_{\Xi\Sigma}\alpha_{\Lambda p}
cos(\theta_{\Sigma\Lambda})cos(\theta_{\Lambda p})]$,
where the angles are determined in the rest frame of each decaying particle.

If one ignores the decay $\siglg$ and simply averages over all directions of
the $\Lambda$ emission, then the net asymmetry seen will be diluted by a
factor of -1/3, assuming that experimental acceptance does not affect the
decay asymmetrically.
It should be noted
that the previous measurement of asymmetry in the decay $\cassg$ did not take
into account this dilution\cite{Teige}.

In our analysis, we consider the 2-dimensional distribution of the cosines
described above.
We measure the rest frame angles for each event 
and then perform a $\chi^2$ comparison
between the 
distribution seen in data with that for Monte Carlo, using different values
of the asymmetry parameter in the generation.  
We used a 5 bin by 5 bin construction. A row of 5 bins 
is virtually empty due to the cut made on the
$\pi^\circ$ mass so 
we do not use these bins in the $\chi^2$ calculation.   We made
11 separate Monte Carlo samples using different values of the asymmetry
parameter, ranging from $\alpha_{\Xi\Sigma}=0.0$ 
to $\alpha_{\Xi\Sigma}=-1.0$.  
The $\chi^2$ 
between the background subtracted data and the Monte Carlo distributions
was calculated
for each of these cases and the result is shown in Figure 3.  A fit to
this curve with a parabola gives a minimum at $\alpha_{\Xi\Sigma}=-0.63$, 
with a corresponding value of $\chi^2$ of 27.7 for 18 degrees of freedom.
The 1 unit variation of 
$\chi^2$ gives an error on this measurement of $\pm0.08$.

Several tests were done to see if there are 
systematic effects on this result for
the asymmetry parameter.
To test the effect of background on the result for asymmetry, we
repeated the analysis described above without background
subtraction.  The result is $\alpha_{\Xi\Sigma}=-0.66\pm0.08$, which does
not vary from the background subtracted value.  The main difference is
that the $\chi^2$/DOF at the minimum increases from 1.5 to 2.8.
To test whether the beam hole region had an effect on the calculation of
the asymmetry value, we applied the additional requirement that the
pion does not hit the region of the calorimeter between the beam holes.
We then repeated the $\chi^2$ analysis.  The result is 
$\alpha_{\Xi\Sigma}=-0.67\pm0.08$
with the $\chi^2$/DOF at the minimum of 1.5.
We also calculated the $\chi^2$ fit for comparisons with the Monte Carlo
data without HA simulation and found no significant difference.  
We estimate that the systematic error on the asymmetry measurement is
$\pm0.05$, yielding our final result, 
${\alpha_{\Xi\Sigma}=-0.63\pm0.08\pm0.05}$.
We independently verified the sign and magnitude of this asymmetry by 
directly comparing our two oppositely polarized samples of data.

This result is only the second measurement making an 
accurate determination of a WRHD asymmetry.
The value of the decay asymmetry is highly negative, similar
to that seen in the decay $\Sigma^+\rightarrow p\gamma$.
\vspace*{.2 in}

\begin{figure} 
\centerline{\psfig{figure=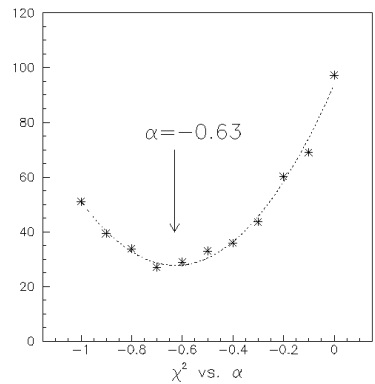,width=6cm}}
\caption{The plot of $\chi^2/DOF$ comparison of background subtracted data
cosine distributions to the Monte Carlo.
The 2nd degree polynomial fit is shown, with the minimum at $\alpha=-0.63$
identified.}
\label{chi2}
\end{figure}

We gratefully acknowledge the support and effort of the Fermilab
staff and the technical staffs of the participating institutions for
their vital contributions.  This work was supported in part by the U.S. 
Department of Energy, The National Science Foundation and The Ministry of
Education and Science of Japan. 
In addition, A.R.B., E.B. and S.V.S. 
acknowledge support from the NYI program of the NSF; A.R.B. and E.B. from 
the Alfred P. Sloan Foundation; E.B. from the OJI program of the DOE; 
K.H., T.N. and M.S. from the Japan Society for the Promotion of
Science.  P.S.S. acknowledges receipt of a Grainger Fellowship.

\end{document}